\documentclass[journal]{IEEEtran}
\usepackage{cite}
\usepackage[pdftex]{graphicx}
\usepackage{epstopdf}
\usepackage[cmex10]{amsmath}
\usepackage{algorithmic}
\usepackage{amssymb}
\usepackage{color}
\usepackage{bm}
\usepackage[keeplastbox]{flushend}
\newcommand{\cc}[1]{#1}
\newcommand{\dd}[1]{#1}
\newcommand{\ee}[1]{#1}

\begin{document}

\title{Weighted Selection Combinings for Differential Decode-and-Forward Cooperative Networks }

\author{Yi~Lou,
		Qi-Yue~Yu,~\IEEEmembership{Member,~IEEE},
		Julian~Cheng,~\IEEEmembership{Senior~Member,~IEEE} and 
        Hong-Lin~Zhao,~\IEEEmembership{Member,~IEEE}
        
\thanks{Y. Lou, Q.-Y. Yue and H.-L. Zhao are with the Department of Communication Engineering, Harbin Institute of Technology, Harbin 150001, China (e-mail: hitbourne@gmail.com).}
\thanks{J. Cheng is with the School of Engineering, The University
of British Columbia, Kelowna, BC V1X 1V7, Canada (e-mail:julian.cheng@ubc.ca)}
\thanks{H.-L. Zhao is the corresponding author.}}

\maketitle

\begin{abstract}
Two weighted selection combining (WSC) schemes are proposed for a differential decode-and-forward relaying system in Rayleigh fading channels. Compared to the conventional selection combining scheme, the decision variable of the relay link is multiplied by a scale factor to combat the error propagation phenomenon. \cc{Average bit-error rate (ABER)} expressions of the two proposed WSC schemes are derived in closed-form and verified by simulation results. \ee{ For the second WSC scheme, asymptotic ABER expression and diversity order are derived to gain more insight into this scheme.} \cc{Moreover, it is demonstrated that both WSC schemes can overcome the extra noise amplification induced by the link adaptive relaying scheme. The first WSC scheme is slightly inferior to the second one, which has a higher complexity. Both proposed WSC schemes outperform the conventional selection combining scheme.}

\end{abstract}

\section{Introduction}

\IEEEPARstart{T}{he} main performance degradation in a decode-and-forward system \cc{lies} in the error propagation due to the possible detection errors \cc{at} the relay node. To cope with this problem, various approaches have been investigated for coherent systems\cite{Wang2008Smart,Zhao2014smart,Wang2007high,Kim2015low,Lu2015End}. For an example, in the link adaptive relaying (LAR) technique \cite{Wang2008Smart}, the relay node adapts to reduce the transmission power according to the instantaneous \cc{channel state information (CSI) }of both hops of the relay system.

Most of the existing works on cooperative networks assume full knowledge of CSI for coherent detection at relay and destination nodes, and an overhead of frequent channel estimation is required at both nodes. The three-node differential decode-and-forward (D-DF) has been studied in \cite{Zhao2007Differential,Zhu2010Differential,wang2006link,Himsoon2007Differential,Himsoon2008Differential,Ikki2009Performance,Cui2009Differential,Song2009A,Yuan2010Differential,Gao2011Performance,Bhatnagar2012Decode}. In \cite{Zhao2007Differential,Zhu2010Differential}, the authors derived the maximum likelihood (ML) detector for the D-DF relaying sytems and the low-complexity piecewise linear detector which closely matches the performance of ML detector.  In \cite{wang2006link}, the authors extended the line of work in~\cite{Wang2008Smart} to the cooperative system using differential modulation. However, no error analysis is presented. For the differential modulation, LAR scheme is not suitable since lowering the transmission power at the relay introduces additional performance degradation, which is caused by increasing the noise component in the decision variable of the relay link.

In \cite{Zhao2007Differential,Zhu2010Differential,wang2006link,Himsoon2007Differential,Himsoon2008Differential,Ikki2009Performance,Cui2009Differential,Song2009A,Yuan2010Differential,Gao2011Performance,Bhatnagar2012Decode}, the received signals of all the links are processed at the destination to achieve the diversity gain. In selection combining (SC), only one of the diversity branches is required to be processed. In \cite{Avendi2013Selection}, the authors analyzed the performance of SC in differential amplify-and-forward systems. However, due to the error propagation, the achieved diversity order of SC in the D-DF systems is only one.

This letter focuses on the D-DF system with a single relay over slow  fading channels. To alleviate the error propagation problem, two weighted selection combining (WSC) schemes are proposed with different complexity requirements. Both schemes avoid the additional performance deterioration in the LAR scheme. Closed-form average bit-error rate (ABER) expressions are derived for both WSC schemes and asymptotic ABER expression and diversity order are derived for the second WSC scheme to gain more insight into the scheme. Compared to the  SC scheme, both proposed schemes can provide significant improvements in error performance.

\section{System Model and Weighted Selection Combining}
Consider a D-DF system, where a source node S communicates with a destination node D via a direct link and a two-hop link through a relay node R. All the nodes are equipped with a single antenna and operate in half-duplex mode. 

Before transmission, the \cc{binary phase-shift keying} modulated symbol $d(k) \in\{+1,-1\}$ is encoded differentially as $s(k)=s(k-1) d(k)$ with $s(0)=1$. In phase I, $s(k)$ is transmitted to R and D from S \cc{with the average symbol power $P_0$}. The corresponding received signals at D and R are
\begin{IEEEeqnarray}{lCl}
y_0(k)&=& \sqrt{P_0}h_0(k) s(k)+n_0(k),\label{y0}\\
y_1(k)&=& \sqrt{P_0}h_1(k) s(k)+n_1(k)\label{y1}
\end{IEEEeqnarray}
\cc{where $h_0$ and $h_1$ denote the channel fading gains of the S-D and S-R channels, respectively. In addition, $n_0$ and $n_1$ denote the additive noise at the corresponding terminals.} The received symbols at R are first differentially decoded by using the detection rule $\hat{d}(k)=\text{sign} \{\text{Re}\{y_1(k)y^*_1(k-1)\}\}$. Then $\hat{d}(k)$ is differentially remodulated into $\hat{s}(k)$, which is then forwarded to D in phase II\@. The signal received at D is given as
\begin{IEEEeqnarray}{C}
y_2(k) = \sqrt{P_0} h_2(k) \hat{s}(k)+n_2(k)\label{y2}
\end{IEEEeqnarray}
\cc{where $h_2$ denotes the channel fading gain of the R-D channel and $n_2$ denotes the additive   noise at D. In \eqref{y0}-\eqref{y2}, $h_i(k), i=0,1,2$, are modeled as complex Gaussian random variable with mean zero and variance $\sigma _i^2$, and $n_i(k)$ are modeled as  $\mathcal{CN}(0,N_0)$.} The instantaneous and average SNR are respectively given as $\gamma_{i}=P_0 |h_i|^2/N_0$ and $\bar{\gamma}_{i}=P_0 {\sigma_i}^2/N_0$.

By using the receiving signals from S in phase I and that from R in phase II, the decision variable for the direct and relay link can be computed as $\xi_{i}=\text{Re}\left\{y_i(k)y^*_i(k-1)\right\}$, $i=0,2$, respectively. In both proposed WSC schemes, these decision variables are combined using the rule
\begin{IEEEeqnarray}{C}\label{xiswitch}
\xi=
\begin{cases}
\xi_{0},      & |\xi_{0}|>\beta|\xi_{2}| \\
\xi_{2},      & |\xi_{0}|<\beta|\xi_{2}|\cc{.}
\end{cases}
\end{IEEEeqnarray}

\cc{Unlike the conventional SC scheme, where the link with a larger magnitude of the corresponding decision variable is selected for non-coherent detection, in the WSC schemes, a novel weight factor $\beta$ is introduced in the selection process with the goal to alleviate the performance degradation caused by the error propagation.} Finally, the transmitted information is detected as $\hat{d}(k)=\text{sign} \left\{\xi \right\}$.

In the following, we propose \cc{two new} WSC algorithms by defining two different types of the weight factor $\beta$. 
\subsection{WSC1 Scheme}
The closed-form ABER expression of the D-DF system is first derived as a function of $\beta$ and \cc{the average SNR of all the links.} Then, the optimal weight factor $\beta_{\text{opt}}$ \cc{is the one that minimizes ABER}. WSC1 is defined as the D-DF system that assumes $\beta=\beta_{opt}$ in~\eqref{xiswitch}.  It is easy to show that the conventional SC scheme \cc{is a} special case of proposed WSC1 scheme with $\beta=1$. 
\subsection{WSC2 Scheme}
In WSC2, by exploiting the instantaneous SNR of S-R link $\gamma_{\text{sr}}$ and taking into account the decoding error at R, an adaptive version of $\beta$ can be given as
\begin{IEEEeqnarray}{C}\label{weightwsc2}
\beta=
\begin{cases} \frac{\gamma _1}{\bar{\gamma }_2}, & \gamma _1<\bar{\gamma }_2 \\
 1, & \gamma _1\geq \bar{\gamma }_2\cc{.}
\end{cases}
\end{IEEEeqnarray}

It should be noted that, although the weight factor \cc{$\beta$} in \eqref{weightwsc2} is given in the same form as the power scale factor \cc{$\beta_L$} used at R in the LAR scheme \cite{wang2006link,Wang2008Smart}, the weight factor is more suitable for the differential modulation as explained in the next paragraph. 

The decision variable of the relay link using LAR scheme can be derived as
\begin{IEEEeqnarray*}{lCl}
\xi_{L}&=&\text{Re}\{y_L(k) y_L^*(k-1)\}\\
&=&\beta_L P_0 |h_2(k)|^2 d(k)+\sqrt{\beta_L}\sqrt{P_0}\hat{s}(k)n_L(k)\\
&&+\text{Re}\{n_2(k)n_2^*(k-1)\}\quad \IEEEyesnumber
\end{IEEEeqnarray*}
where $y_L(k)=\sqrt{\beta_L} \sqrt{P_0} h_2(k) \hat{s}(k)+n_2(k)$ and $n_L(k)=|h_2(k)|\text{Re}\{n_2^*(k-1) + n_2(k)\}$ \cc{are the received signal of the relay link at D in the LAR scheme and the corresponding one-order noise term.} \cc{In addition, we have utilized the fact that $h_2(k)= h_2(k-1)$ in the block fading and the rotational invariance of complex Gaussian variable.} As $\beta_L\in(0,+1]$, $\beta_L<\sqrt{\beta_L}$.  Hence, LAR scheme will increase \cc{ the dominating noise component} in the decision variable of the relay link and \cc{cause} an additional performance degradation in the decision process. This will be verified in the simulations.

For the blocking fading channels, $\gamma _1$ can be approximated by R as $\gamma _1=E\{\boldsymbol{y}_1^H \boldsymbol{y}_1\}/(L N_0)-1\approx\|\boldsymbol{y}_1\|^2/(L N_0)-1$, where $\|\boldsymbol{y}_1\|^2=\boldsymbol{y}_1^H \boldsymbol{y}_1$, \cc{$\boldsymbol{y}_1=[y_1(1), \dots, y_1(L)]^T$} \cc{and $L$ is the block length.} Then $\gamma _1$ can be forwarded to D only once per frame.

\section{Performance Analysis}
In this section, we develop closed-form ABER expressions of the two proposed WSC schemes for the D-DF system in closed-form. 

Without loss of generality, we assume $d(k)=1$. The ABER of both proposed schemes \cc{can be expressed in a unified form as}
\begin{IEEEeqnarray*}{lCl}\label{petotal}
P_e&=&\Pr \left(\left| \xi _0\right|  >\beta  \left| \xi _2\right|  ,\xi _0<0\right)+\Pr \left(\left| \xi _0\right| <\beta  \left| \xi _2\right|  ,\xi _2<0\right)\\
&=&\underbrace{\Pr \left(\left| \xi _0\right|  > \left| \xi _w\right|  ,\xi _0<0\right)}_{P_{e1}}+\underbrace{\Pr \left(\left| \xi _0\right|  < \left|\xi _w\right|  , \xi _w<0\right)}_{P_{e2}}\quad\IEEEyesnumber
\end{IEEEeqnarray*}
where $ \xi _w=\beta \xi_2 $. The first term of in \eqref{petotal} can be calculated as
\begin{IEEEeqnarray*}{lCl}\label{pe1}
P_{e1}&=&\Pr \left(\left| \xi _w\right| +\xi _0<0\right)\\
&=&\int _{-\infty }^0\int _0^{-x}f_{\xi _0}(x)f_{\left| \xi _w\right| }(r)drdx\\
&=&\int _{-\infty }^0f_{\xi _0}(x)\left(F_{\left| \xi _w\right| }(-x)-F_{\left| \xi _w\right| }(0)\right)dx.\IEEEyesnumber
\end{IEEEeqnarray*}

By following the \cc{approach similar} to \eqref{pe1}, the second term of in \eqref{petotal} is given as
\begin{IEEEeqnarray*}{C}\label{pe2}
P_{e2}=\int _{-\infty }^0f_{\xi _w}(x)\left(F_{\left| \xi _0\right| }(-x)-F_{\left| \xi _0\right| }(0)\right)dx.\IEEEyesnumber
\end{IEEEeqnarray*}
where
\begin{IEEEeqnarray}{C}\label{absxi}
F_{\left|\xi _i\right|}(x)=F_{\xi _i }(x)-F_{\xi _i }(-x),\quad i={0,w}.
\end{IEEEeqnarray}

Define \(A_i=\frac{\bar{\gamma }_i}{\bar{\gamma }_i+1} \left|y_i(k-1)\right|^2\) and \(B_i=\frac{N_0}{2}\left(\frac{2\bar{\gamma }_i+1}{\bar{\gamma }_i+1}\right)\left|y_i(k-1)\right|^2\), where $i\in\{0,2\}$. It is easy to show that, conditioned on a given \(y_0(k-1)\), $\xi _0$ is a real \cc{Gaussian} random variable with mean \(A_0\) and variance \(B_0\), i.e., \(\xi _0\sim \mathcal{N}\left(A_0,B_0\right)\)~\cite{kam1991bit}. The corresponding conditional \cc{probability density function (pdf)} of $\xi _0$ is given by
\begin{IEEEeqnarray}{C}\label{xi0Condiy0}
f_{\xi _0}\left(x\left|y_0\right.\right)=\frac{1}{\sqrt{2\pi  B_0}}\text{exp}\left(-\frac{{\left(x-A_0\right)}^2}{2B_0}\right)
\end{IEEEeqnarray}
where the time index \(k-1\) has been dropped for brevity.

From \eqref{y0}, it is straightforward to show that, for a given $s(k)$, $y_0\sim \mathcal{CN}\left(0,N_0\left(\bar{\gamma}_0+1\right)\right)$\cc{; therefore}, \(\left|y_0\right|^2\) follows \cc{an} exponential distribution with parameter $N_0\left(\bar{\gamma }_i+\cc{N_0}\right)$. The corresponding pdf is given by \(f_{\left|y_0\right|^2}(x)=\frac{1}{N_0\left(\bar{\gamma}_0+1\right)}e^{-\frac{x}{N_0\left(\bar{\gamma }_0+1\right)}}\).

To simplify the notation, \cc{we define }\cc{$u_i=1+\bar{\gamma}_i$ and $v_i=1+2\bar{\gamma}_i$, where $i\in\{0,1,2\}$.} The  pdf of $\xi_0$ can be obtained by averaging the conditional pdf in \eqref{xi0Condiy0} with respect to \(f_{\left|y_0\right|^2}(x)\), and it is given by
\begin{IEEEeqnarray}{C}\label{fxi0}
f_{\xi _0}(x)=
\begin{cases}
\frac{1}{u_0} \text{exp}(2x), & x\leq0 \\
\frac{1}{u_0} \text{exp}\left( -\frac{2 x}{v_0} \right), & x\geq0\cc{.}
\end{cases}
\end{IEEEeqnarray}

The corresponding \cc{cumulative distribution function (CDF)} is derived as
\begin{IEEEeqnarray}{C}\label{{xi0CDF}}
F_{\xi _0}(x)=
\begin{cases}
\frac{1}{2 u_0}\text{exp}\left(2 x\right), & x\leq 0 \\
1-\frac{ v_0}{2 u_0}\text{exp}\left( -\frac{2 x}{v_0} \right), & x\geq0\cc{.}
\end{cases}
\end{IEEEeqnarray}

With the help of~\eqref{absxi}, the CDF of \(\left|\xi _0\right|\) is given as
\begin{IEEEeqnarray}{C}\label{Fabsxi0}
F_{\left|\xi _0\right|}(x)=1-\frac{1}{2 u_0}\text{exp}\left( -2x \right)-\frac{v_0}{2 u_0}\text{exp}\left( -\frac{2 x}{v_0}  \right).
\end{IEEEeqnarray}

Furthermore, conditioned on \(y_2(k-1)\) and $\hat{d}(k)$, \(\xi _2\sim \mathcal{N}\left(\hat{d}(k) A_2,\hat{d}^2(k) B_2\right)\). Consequently, conditioned on $\beta$, \(\xi _w=\beta \xi _2  \sim \mathcal{N}\left(\beta \hat{d}(k) A_2,\beta^2 \hat{d}^2(k) B_2\right)\). Due to the potential decoding errors at R, the decision variable of the relay link follows a \cc{Gaussian} mixture distribution with a conditional pdf given by
\begin{IEEEeqnarray*}{lCl}\label{mixdist}
\cc{f_{\xi _w}(x|\beta)}&=&P_1 f_{\xi _w}(x|\hat{d}(k)=-1,\beta)\\
&&+\left(1-P_1\right) f_{\xi _w}(x|\hat{d}(k)=1,\beta)\IEEEyesnumber
\end{IEEEeqnarray*}
in which $P_1=\frac{1}{2}e^{-\gamma _1}$ is the instantaneous decoding error probability at R. It can be noticed from~\eqref{mixdist} that \cc{$f_{\xi _w}(x|\hat{d}(k)=-1,\beta) \sim \mathcal{N}\left(-\beta  A_2,\beta ^2 B_2\right)$ and $f_{\xi _w}(x|\hat{d}(k)=1,\beta) \sim \mathcal{N}\left(\beta  A_2,\beta ^2 B_2\right)$.} Eq.~\eqref{mixdist} can be further simplified to
\begin{IEEEeqnarray*}{C}\label{fxi2}
f_{\xi _w}(x|\beta)=\begin{cases}
  \frac{\text{exp}\left( \frac{2 x}{v_2 \beta }-\gamma _1 \right)+\psi(\gamma_1) \text{exp}\left( \frac{2 x}{\beta } \right) }{2 u_2 \beta },
 & x\leq0 \\
  \frac{\text{exp}\left( -\frac{2 x}{\beta }-\gamma _1 \right)+\psi(\gamma_1) \text{exp}\left( -\frac{2 x}{v_2 \beta } \right) }{2 u_2 \beta },
  & x\geq0
\end{cases}\IEEEyesnumber
\end{IEEEeqnarray*}
where $\psi(\gamma_1)=2-\text{exp}\left( -\gamma _1 \right)$.

The corresponding CDF of $\xi _w$ can be derived as
\begin{IEEEeqnarray*}{lCl}
F_{\xi _w}(x|\beta)=
\begin{cases}
  \frac{\psi(\gamma_1) \text{exp}\left( \frac{2 x}{\beta } \right) +v_2 \text{exp}\left( \frac{2 x}{v_2 \beta }-\gamma _1 \right) }{4 u_2},
 & x\leq 0 \\
 1-\frac{\text{exp}\left( -\frac{2 x}{\beta }-\gamma _1 \right)+v_2\psi(\gamma_1) \text{exp}\left( -\frac{2 x}{v_2 \beta } \right) }{4 u_2},
 & x\geq 0.
 \end{cases}\\\IEEEyesnumber
\end{IEEEeqnarray*}

With the help of~\eqref{absxi}, the CDF of \(\left|\xi _w\right|\) is given as
\begin{IEEEeqnarray}{C}\label{Fabsxi2}
F_{\left|\xi _w\right|}(x|\beta)=1-\frac{\text{exp}\left( -\frac{2 x}{\beta } \right)+v_2 \text{exp}\left( -\frac{2 x}{v_2 \beta } \right) }{2 u_2}.
\end{IEEEeqnarray}

Substituting~\eqref{fxi0} and~\eqref{Fabsxi2} into~\eqref{pe1}, the integral in~\eqref{pe1} can be solved in closed-form as
\begin{IEEEeqnarray*}{lCl}
P_{e1}^{WSC1}&=&\frac{u_2+v_2 \beta }{2 u_0 u_2 (1+\beta)  (1+v_2 \beta)}.
\end{IEEEeqnarray*}

Substituting \eqref{Fabsxi0} and \eqref{fxi2} into \eqref{pe2} and then taking the expectation with respect to the exponential distribution of $\gamma_1$, $P_{e2}^{WSC1}$ can be solved and expressed as
\begin{IEEEeqnarray*}{lCl}
P_{e2}^{WSC1}&=&\frac{\beta  \left(I_1+I_2\right)}{2 u_0 u_1 u_2 (1+\beta ) \left(v_0+\beta \right) \left(1+v_2 \beta\right) \left(v_0+v_2 \beta \right)}
\end{IEEEeqnarray*}
\cc{where}
\begin{IEEEeqnarray*}{lCl}
I_1&=&-(1-u_2) \big(2 u_2 v_0^2-u_0 (1-2 u_2-4 u_2^2) v_0 \beta\\
 &&\:+4 u_0^2 u_2 v_2 \beta ^2+u_0 v_2^2 \beta ^3\big)
 \end{IEEEeqnarray*}
 and
 \begin{IEEEeqnarray*}{lCl}
 I_2&=&u_1 \Big\{v_2 \beta ^2 \big[1-2 u_2-u_0 (2-2 u_0-4 u_2-v_2 \beta )\big]\\
 &&\:+v_0^2-u_0 (1-4 u_2) v_0 \beta \Big\}.
\end{IEEEeqnarray*}

Hence, the ABER using the WSC1 scheme is obtained in closed-form as
\begin{IEEEeqnarray}{C}\label{berwsc1}
P_{e}^{WSC1}=P_{e1}^{WSC1}+P_{e2}^{WSC1}.
\end{IEEEeqnarray}

To find the optimal value, $\beta_\text{opt}$ of the weight factor that minimizes the ABER in~\eqref{berwsc1}, \ee{a numerical method can be adopted\cite{Selvaraj2010Scaled,Selvaraj2011Single,Cui2013Weighted,Cui2014Weighted}}. \cc{we differentiate $P_{e}^{WSC1}$ in~\eqref{berwsc1}} with respect to \(\beta\), set the resulting derivation to zero, and solve for $\beta_\text{opt}$ numerically.


To derive the closed-form ABER expression for WSC2, we can follow the similar approach used in WSC1 scheme except that \dd{$\beta$ is defined as a piecewise function with intervals that are determined by the instantaneous SNR $\gamma_1$ and average SNR $\bar{\gamma}_2$, as can be seen in~\eqref{weightwsc2}.} The ABER using the WSC2 scheme is obtained in closed-form as $P_{e}^{WSC2}=P_{e1}^{WSC2}+P_{e2}^{WSC2}$. The first term of the ABER expression \cc{$P_{e1}^{WSC2}$} can be derived as
\begin{IEEEeqnarray*}{lCl}
P_{e1}^{WSC2}&=&\int _{-\infty }^0f_{\xi _0}(x)\left(F_{\left| \xi _w\right| }(-x|\beta )-F_{\left| \xi _w\right| }(0|\beta )\right)dx\\
&=&L_1+L_2\IEEEyesnumber
\end{IEEEeqnarray*}
where
\begin{IEEEeqnarray*}{lCl}
L_1&=&\int _{-\infty }^0\int _0^{\bar{\gamma }_2}f_{\xi _0}(x)(F_{| \xi _w| }(-x|\gamma _1)-F_{|\xi _w| }(0|\gamma _1))\\
&&\:\times f_{\gamma _1}(\gamma _1)d\gamma _1dx\\
&=&\frac{\phi  }{4 u_0 u_2} \text{exp}\left( \frac{\phi }{v_2} \right) \cc{\Bigg\{} \text{exp}\left( -\frac{2 (1-u_2) \phi }{v_2} \right) \Big(E_1\left(\phi \right)\\
&&-E_1\left(2\phi \right)\Big)+E_1\left(\frac{\phi }{v_2}\right)-E_1\left(\frac{2 u_2 \phi }{v_2}\right)\cc{\Bigg\}}
\end{IEEEeqnarray*}
and
\begin{IEEEeqnarray*}{lCl}
L_2&=&\int _{-\infty }^0\int_{\bar{\gamma }_2}^{\infty }f_{\xi _0}(x)\left(F_{\left| \xi _w\right| }(-x)-F_{\left| \xi _w\right| }(0)\right)dx\\
&=&\frac{ \left(3 u_2-1\right)}{8 u_0 u_2^2}\text{exp}\left( -\phi \right)
\end{IEEEeqnarray*}
and where $\phi=\bar{\gamma}_2/\bar{\gamma}_1$, and $E_1(\cdot)$ denotes the exponential integral function. Note that $L_1$ is obtained by applying the result from \cite[eq (3.352.1)]{Gradshteyn2007Table} and performing some simplification. 

The second term of the ABER expression $P_{e2}^{WSC2}$ is given by
\begin{IEEEeqnarray*}{lCl}
P_{e2}^{WSC2}&=&\int _{-\infty }^0f_{\xi _w}(x)\left(F_{\left| \xi _0\right| }(-x|\beta )-F_{\left| \xi _0\right| }(0|\beta )\right)dx\\
&=&K_1+K_2\IEEEyesnumber
\end{IEEEeqnarray*}
where
\begin{IEEEeqnarray*}{lCl}
K_1&=&\int _{-\infty }^0\int _0^{\bar{\gamma }_2}f_{\xi _w}(x)(F_{| \xi _0| }(-x|\gamma _1)-F_{|\xi _0| }(0|\gamma _1))\\
&&\:\times f_{\gamma _1}(\gamma _1)d\gamma _1dx\\
&=&-\frac{\text{exp}\left( -(1+u_1) \phi \right)}{8 u_0 (1-u_1) u_1 u_2}\cc{\Bigg\{}-4 \text{exp}\left( \phi \right) u_0 (1-u_1) \Big(1- u_1\\
&&\times \text{exp}\left( u_2-1 \right) -\text{exp}\left( u_1 \phi \right) (1-u_1-u_2)-u_2\Big)\\
&&+\Xi\left(1,\frac{2(1+u_1) u_2-1}{v_2},\frac{u_1 u_2}{v_2},\frac{u_1}{v_2},\frac{u_1 (u_0+u_2-1)}{u_0 v_2}\right)\\
&&+\Xi\left(2,2+u_1,1,1,1\right)+\Xi\left(-1,1+2u_1,u_1,u_1,u_1\right)\cc{\Bigg\}}
\end{IEEEeqnarray*}
\cc{and}
\begin{IEEEeqnarray*}{lCl}
K_2&=&\int _{-\infty }^0\int_{\bar{\gamma }_2}^{\infty }f_{\xi _w}(x)\left(F_{\left| \xi _0\right| }(-x)-F_{\left| \xi _0\right| }(0)\right)dx\\
&=&\frac{\left(J_1+J_2\right)}{16 u_0^2 u_1 \left(1-u_0-u_2\right) u_2^2} \text{exp}\left( -1-u_1 \phi \right)\IEEEyesnumber
\end{IEEEeqnarray*}
\cc{and where}
\begin{IEEEeqnarray*}{lCl}
J_1&=&e \left(1-u_2\right) \Big\{u_2+u_0 \big(1-7 u_2-u_0 \left(1-4 u_2-8 u_2^2\right)\big)\Big\},\\
J_2&=&2 e^{u_2} \left(3 u_0-1\right) u_1 \left(1-u_0-u_2\right) u_2,
\end{IEEEeqnarray*}
and $\Xi\left(x_1,x_2,x_3,x_4,x_5\right)\cc{\triangleq}\text{exp}( x_2 \phi )  u_1 (u_2-1)  x_1 \times \big(E_1(2 \phi  x_3 )-E_1(\phi  x_4)-\text{exp}( 2 (u_0-1) \phi x_4 ) v_0^2 (E_1(v_0 \phi  x_4)-E_1(2 u_0 \phi  x_5)))$.

We examine the asymptotic performance to gain more insight into the above ABER $P_{e}^{WSC2}$ For this purpose, consider the symmetric scenario of with $\sigma_0^2=\sigma_1^2=\sigma_2^2=1$. Using the McLaurin series representation for the exponential and the exponential integral function (\cite[eq (6.6.2)]{Olver2010NIST}) in $P_{e}^{WSC2}$, and considering only the first two order terms. After some simplification, $P_{e}^{WSC2}$ can be expressed as
\begin{IEEEeqnarray*}{lCl}\label{asymp}
P_{e}^{WSC2}&\approx&\frac{1}{P_0^4+P_0^5+P_0^6+P_0^7+P_0^8}\Big\{0.03-0.09 P_0+P_0^2\\
&&\times (-0.16-0.03 \ln(P_0))+P_0^3 (1.75+0.06 \ln(P_0))\\
&&+P_0^4 (4.53+0.47\ln(P_0))+P_0^5 (3.84+0.63 \ln(P_0))\\
&&+P_0^6 (1.11+0.25 \ln(P_0)\Big\}.\IEEEyesnumber
\end{IEEEeqnarray*}

From \eqref{asymp}, it is easy to show that when $P_0\to \infty$, the diversity order $d$ of the WSC2 scheme is
\begin{IEEEeqnarray*}{lCl}\label{diversityorder}
d=-\lim_{P_0\to\infty }  \frac{\ln (P_e^{\text{WSC2}})}{\ln(P_0)}\approx 2-\frac{\ln(0.25\ln (P_0)) }{\ln(P_0)} \approx 2.
\end{IEEEeqnarray*}

\section{Simulation}
In this section, numerical results are provided to verify the analysis. All links are subject to independent Rayleigh block fading. The block length for simulation is $L = 256$ bits. The number of the blocks for simulation is $10^7$.

\cc{Figure 1} shows the ABER of WSC1 as a function of weight factor $\beta$ with $P_0/N_0=30$ dB. It can be observed that the ABER \cc{reaches} the minimum value when $\beta=\beta_\text{opt}$ for each scenario. Meanwhile, with increasing $\sigma_1^2$, the optimal weight factor increases and \cc{approaches 1}. On the contrary, the optimal weight factor decreases and tends to zero with increasing $\sigma_2^2$. \cc{Intuitively}, the decision variable of the relay link is more reliable with a better channel quality of the S-R link due to the decreasing decoding error propagation at R. In this situation, a larger $\beta$ value will increase the probability of $\xi_{2}$ being selected, and hence WSC1 achieves more performance improvement from the combining gains. Same explanation \cc{also applies} to the case of strong R-D link where a lower average R-D channel gain can alleviate the error propagation.

\cc{Figure 2} verifies the tightness of the derived ABER expressions and asymptotic ABER expression in \eqref{asymp} in the symmetric channel with $\sigma _0^2=\sigma _1^2=\sigma _2^2=1$. The conventional SC, LAR in \cite{wang2006link}, and ML detector in\cite{Zhao2007Differential,Zhu2010Differential} are also plotted as benchmarks. \cc{As a special case}, the ABER performance of the SC scheme can be \cc{obtained} by setting $\beta=1$ in \eqref{berwsc1}. It is seen that the theoretical results match the simulation curves \cc{perfectly} for the two proposed schemes. Moreover, the asymptotic ABER expression in \eqref{asymp} is tight for high $P_0/N_0$. Due to the error propagation, the diversity order of SC scheme is unity. WSC2 can achieve the same performance as that of ML detector. However, the ML detector has exponential complexity and requires non-linear processing. As discussed in Section II, the noise amplification \cc{introduced} by the LAR factor at R degrades the system error performance. On the other side, the WSC1 scheme is also attractive as it strikes a trade-off between performance and complexity.

\begin{figure}[!t]
\centering
\includegraphics[width=3.45in]{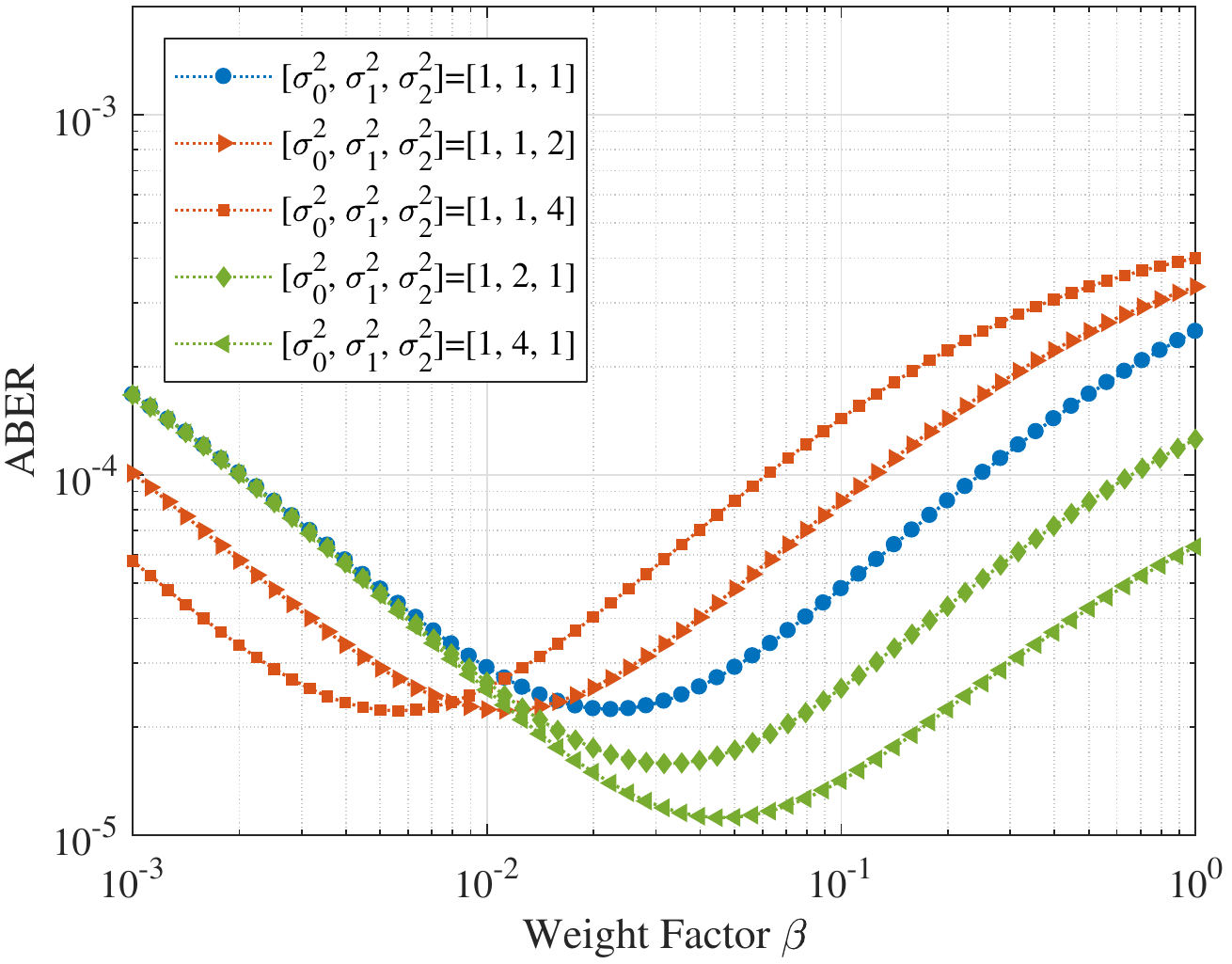}
\caption{ABER performance of the WSC1 scheme versus weight factor $\beta$ for $P_0/N_0=30$ dB .}
\label{berversusbeta}
\end{figure}

\begin{figure}[!t]
\centering
\includegraphics[width=3.45in]{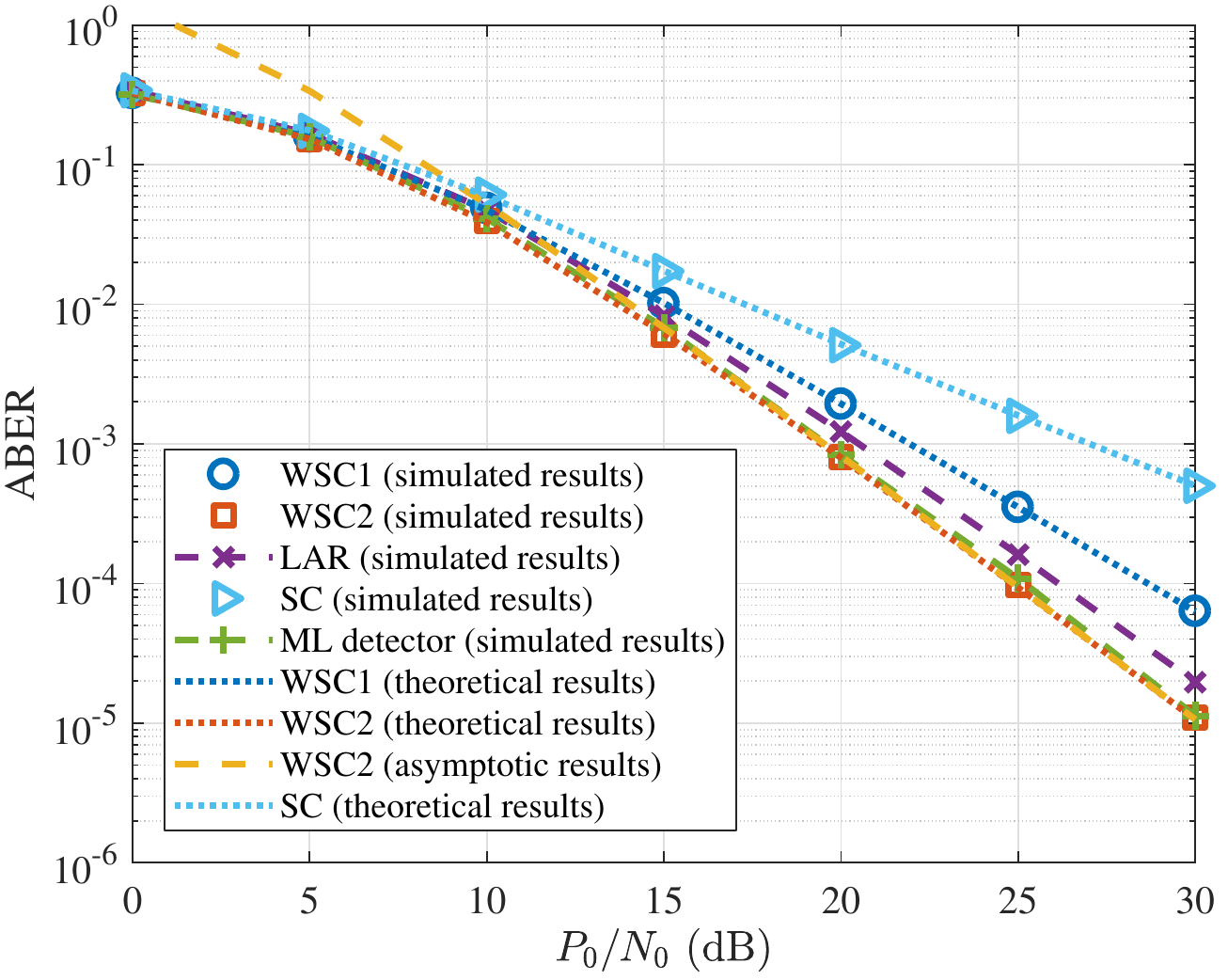}
\caption{ABER performance comparison of the D-DF system using SC, LAR, WSC1, and WSC2 schemes in the symmetric channel with $\sigma _0^2=\sigma _1^2=\sigma _2^2=1$.}
\label{ber}
\end{figure}

\section{Conclusion}
Two weighted selection combining schemes \cc{were} proposed for a differential decode-and-forward relaying system. The exact \cc{bit-error rate} expressions  for both schemes are derived in closed-form. The proposed schemes show significant improvements in error performance when compared to the conventional selection combining scheme, and moreover, the additional performance degradation \cc{introduced} by the noise amplification in the LAR scheme is also avoided.

\bibliographystyle{IEEEtran}

\bibliography{IEEEabrv,mybib}
\end{document}